\documentclass[10pt,sigconf,nonacm]{acmart}
\AtBeginDocument{%
  \providecommand\BibTeX{{%
    Bib\TeX}}}
\usepackage{pifont}
\usepackage{enumerate}
\usepackage{graphicx}
\usepackage{subcaption}
\usepackage{soul}
\usepackage[american]{babel}
\usepackage[capitalise,nameinlink,noabbrev]{cleveref}
\usepackage{listings}
\newcommand{\ourss}{\texttt{PrivateHub}\xspace}
\usepackage{xcolor}
\usepackage[most]{tcolorbox}
\sethlcolor{yellow}
\soulregister{\ourss}{0}
\soulregister{\cite}{1}
\soulregister{\ref}{1}
\soulregister{\Cref}{1}
\soulregister{\cref}{1}
\soulregister{\textsuperscript}{1}
\soulregister{\textcopyright}{0}
\soulregister{\emph}{1}
\soulregister{\textbf}{1}
\soulregister{\textit}{1}

\newtcolorbox{newtext}{colback=yellow!35,boxrule=0pt,arc=0pt,outer arc=0pt,
  left=2pt,right=2pt,top=1pt,bottom=1pt,breakable,enhanced jigsaw,boxsep=1pt}
\usepackage{colortbl} 
\usepackage{xcolor} 
\usepackage{booktabs} 
\usepackage{graphicx}
\usepackage{wrapfig}
\usepackage{epsfig}
\usepackage{caption}
\usepackage{multirow}
\usepackage{stfloats}
\usepackage{balance}
\usepackage{tikzscale}
\usepackage{tikz}
\usepackage{siunitx}
\usepackage{mathtools}
\usetikzlibrary{fit,positioning}
\usepackage{filecontents}
\usepackage{lipsum}
\usepackage[super]{nth}
\usepackage{hhline}
\usepackage{multicol}
\usepackage{multirow}
\usepackage{makecell}
\usepackage{tabularx}
\usepackage{listings}
\usepackage{float}
\usepackage{enumerate}
\usepackage{pifont}
\usepackage{tabularx}     
\usepackage{textcomp}     
\usepackage{threeparttable} 
\usepackage{url}          
\usepackage[usenames,dvipsnames,svgnames]{xcolor} 
\usepackage{xspace}       
\usepackage[toc,page]{appendix}
\usepackage{hhline}
\usepackage{array}
\usepackage{booktabs}

\usepackage{setspace}
\usepackage{algorithmic}

\usepackage{bbding}

\lstdefinelanguage{JavaScript}{
  keywords={break, case, catch, continue, debugger, default, delete, do, else, finally, for, function, if, in, instanceof, new, return, switch, this, throw, try, typeof, var, void, while, with},
  morecomment=[l]{//},
  morecomment=[s]{/*}{*/},
  morestring=[b]',
  morestring=[b]",
  sensitive=true
}

\def\BibTeX{{\rm B\kern-.05em{\sc i\kern-.025em b}\kern-.08emT\kern-.1667em\lower.7ex\hbox{E}\kern-.125emX}}
    
\setcopyright{acmcopyright}
\copyrightyear{2027}
\acmYear{2027}
\acmDOI{XXXXXXX.XXXXXXX}

\acmConference[SenSys '27]{The ACM/IEEE International Conference on Embedded Artificial Intelligence and Sensing Systems}{[Month Day--Day], 2027}{New York, NY, USA}

\acmPrice{15.00}
\acmISBN{978-1-4503-XXXX-X/18/06}

\begin{document}

\title
{\ourss: Contrastive Diffusion Model for Private Sensor-Intensive Environment Data Generation}

\renewcommand{\shortauthors}{Gao et al.}

\author{Jiechao Gao}
\authornote{Work done while at the University of Virginia.}
\affiliation{%
  \institution{Stanford University}
  \city{Stanford}
  \state{California}
  \country{USA}}
\email{jiechao@stanford.edu}

\author{Yuandong Pan}
\affiliation{%
  \institution{Stanford University}
  \city{Stanford}
  \state{California}
  \country{USA}}
\email{ydpan@stanford.edu}

\author{Jie Wang}
\affiliation{%
  \institution{Stanford University}
  \city{Stanford}
  \state{California}
  \country{USA}}
\email{jiewang@stanford.edu}

\author{Michael Lepech}
\affiliation{%
  \institution{Stanford University}
  \city{Stanford}
  \state{California}
  \country{USA}}
\email{mlepech@stanford.edu}

\author{Bradford Campbell}
\affiliation{%
  \institution{University of Virginia}
  \city{Charlottesville}
  \state{Virginia}
  \country{USA}}
\email{bradjc@virginia.edu}

\begin{abstract}
Sensor-intensive environments enable many intelligent services by inferring user applications from heterogeneous data streams. However, not all applications should be exposed: users want some activities to stay private. This creates a tension between inferring applications for useful services and preventing unwanted inference. Existing approaches such as differential privacy and rule-based filtering protect individual streams but cannot address the privacy risk from cross-sensor inference.

We introduce \ourss, which uses contrastive learning within a diffusion model to generate synthetic multi-sensor streams that keep non-private applications detectable while concealing private ones. \ourss~has two stages: App-Conditioned Pre-training (ACP), which conditions the model on multi-sensor data with application embeddings, and App-Aware Fine-tuning (AAF), which separates private from non-private data via contrastive learning. We also define a threat model for the multi-sensor sharing setting. Experiments on three real-world multi-sensor datasets show \ourss~lowers private-application accuracy by 40 to 50\% without hurting non-private performance, and stays robust when the attacker retrains on the synthetic data.
\end{abstract}

%
%

%

\settopmatter{printfolios=true}
\maketitle

\section{Introduction}

Internet of Things (IoT) connects many sensors to enable intelligent applications such as traffic forecasting~\cite{jiang2022big}, health monitoring~\mbox{\cite{sujith2022systematic}}, and human activity recognition (HAR)~\cite{anguita2013public}. This rich sensing also brings sensor-data privacy challenges. Sensor-data privacy is often viewed from the perspective of a single stream: for example, bodily data is treated as sensitive~\mbox{\cite{sun2010privacy}}. However, in sensor-intensive environments, fusing multiple streams can reveal sensitive information even when each individual stream looks benign. In a smart space with thermostats and off-the-shelf Doppler, temperature, and humidity sensors~\mbox{\cite{wu2021smart}}, the streams can infer many indoor activities. Some are non-private (e.g., standing, sitting), but others are highly private (e.g., Zoom calls). Identifying non-private activities while concealing private ones is a practical and pressing challenge, and it grows more important as deployed sensors multiply.

\begin{figure}
\centering

      \begin{subfigure}[h]{1\linewidth}
      \centering
      \includegraphics[width=\columnwidth, height=3cm, keepaspectratio= true]{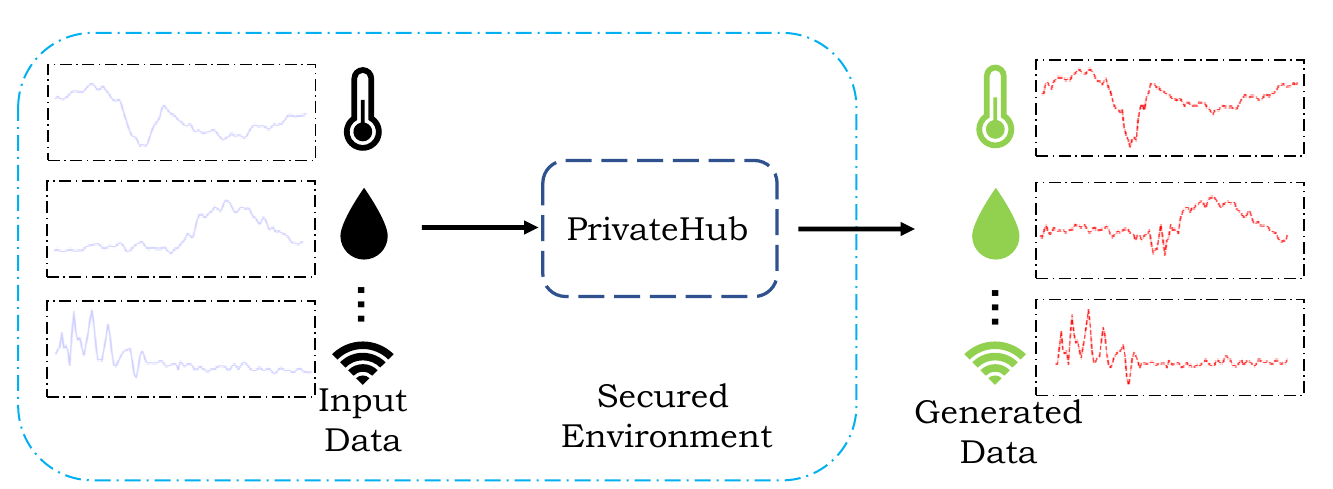}
      \label{fig:down_the_stairs}
      \end{subfigure}
      \vspace{-7pt}
    \caption{High-level architecture of \ourss. Raw multi-sensor data streams are processed locally in a secured environment, and the output synthetic streams can then be safely shared with third-party services.}

  \label{fig: run}
\end{figure}

There are three kinds of traditional privacy protection methods for IoT data streams. (1) \textbf{Rule-based privacy frameworks}~\cite{lola2023towards,aljeraisy2021privacy,gheisari2021obpp} govern data handling through policies such as consent, access limits, and encryption. However, they depend on humans to formulate and enforce rules, so they are prone to errors and oversight. More importantly, they treat each stream independently and miss the composite inference risk from sensor fusion, so private activities can still be inferred even when the rules are followed. (2) \textbf{Differential privacy based frameworks}~\cite{husnoo2021differential,gao2022pfed,ghayyur2018iot} add noise under a privacy budget \mbox{$\epsilon$}~\mbox{\cite{dwork2006our}}. This works for a single stream but not for sensor-intensive settings, because adding noise to all streams ignores the connections among sensors and degrades every application, private or not. (3) \textbf{Generative models for data privacy}~\cite{huang2021robust,yao2023privacy,yang2023privacy} generate synthetic data that resembles the original without exposing it. However, VAE~\mbox{\cite{doersch2016tutorial}}, GAN~\mbox{\cite{goodfellow2020generative}}, and conditional GAN~\mbox{\cite{mirza2014conditional}} suffer from posterior and mode collapse. Diffusion models~\mbox{\cite{yang2023diffusion}} are more stable and produce higher-quality samples, but generating data conditionally to hide private applications while keeping non-private ones detectable is still beyond current diffusion models.

A natural question is whether a simpler strategy could solve this problem without a generative model. For example, one could replay previously recorded non-private segments in place of private ones, or add noise to every sensor stream until private activities become hard to detect. We study both ideas as baselines in Section~\ref{sec:rq1}. Neither works well. Replaying segments breaks the natural timing and the correlations across sensors, so the accuracy of non-private applications drops sharply. Adding noise has to be applied to all streams at once, which also harms useful tasks such as temperature and CO$_2$ prediction, and it only changes the surface of the data, so an attacker can adapt to it by retraining. The real difficulty is therefore not to block inference, but to mislead it selectively: keep non-private activities readable while hiding private ones and keeping the data meaningful. This is what motivates a targeted generative approach.

Therefore, protecting application privacy in sensor-intensive environments requires a solution that satisfies three key requirements: (1) Accurate identification of non-private applications; (2) Successfully concealing private applications; (3) the sensor datastreams are meaningful. (e.g., smart thermostats data can still perform temperature prediction.)

In this paper, we propose \ourss, a synthetic data-stream generation method that uses contrastive learning within a diffusion model to conditionally generate streams that identify non-private applications while concealing private ones. Figure~\mbox{\ref{fig: run}} shows its high-level architecture: users feed their multi-sensor streams into the framework in a secure environment, and the output synthetic streams can detect non-private applications but fail to detect the private ones labeled by the data owner.

\ourss~has two stages. App-Conditioned Pre-training (ACP) pre-trains a diffusion model on multi-sensor streams with application category embeddings. App-Aware Fine-tuning (AAF) then uses contrastive learning, with an off-the-shelf classifier extracting features, to push generated data closer to non-private and away from private. We evaluate on real-world datasets from smart homes, a smart office, and an in-the-wild smartphone-and-smartwatch setting, showing that \ourss~mitigates the sensor-intensive environment privacy challenge while outperforming baselines.


\section{Motivation}
\label{sec:motivation}

\subsection{Privacy Risks in Sensor-Intensive Environments}

\begin{figure}[t!]
\centering
\includegraphics[width=0.8\linewidth, trim=0 0 0 0]{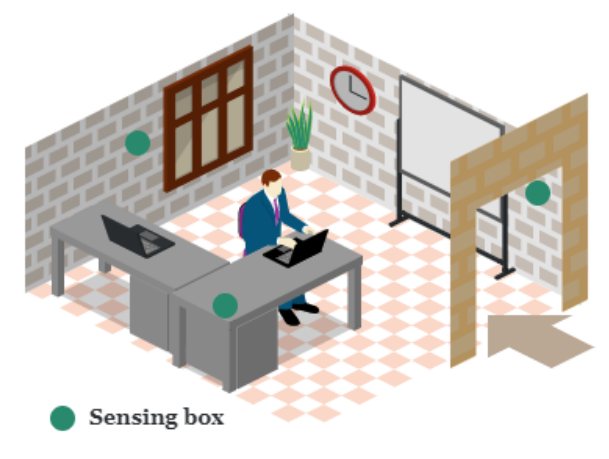}
\caption{Example of sensor data collection.
}
\label{fig: uva}
\end{figure}

Data privacy work usually focuses on the confidentiality of single streams and overlooks sensor fusion. Yet even when individual streams look benign, their aggregation can reveal sensitive behavior: private activities become inferable once temperature, audio, motion, and environmental sensors are analyzed together. As sensing becomes ubiquitous, this fusion risk is increasingly important. To show this is a real and prevalent problem, we ran a preliminary study in two office rooms. We placed AWAIR Omni\mbox{\textsuperscript{\textcopyright}} sensors (Figure~\mbox{\ref{fig: uva}}) that collect indoor environmental quality factors such as \mbox{$\mathrm{CO}_2$}, PM2.5, illuminance, sound pressure level, humidity, and temperature,as shown in Table~\ref{Awair_characteristics}. Five participants each performed four typical office activities: (1) typing, (2) Zoom\mbox{\textsuperscript{\textcopyright}} calls, (3) writing on a whiteboard, and (4) standing while drinking. The first three are commonly seen as private, while standing while drinking is a benign, health-related activity~\mbox{\cite{5555}}. Ideally the streams should detect only the benign activity and not the private ones.

\begin{figure}[t!]
\centering
\scalebox{0.7}{\includegraphics[width=1.0\linewidth, trim=0 0 0 0]{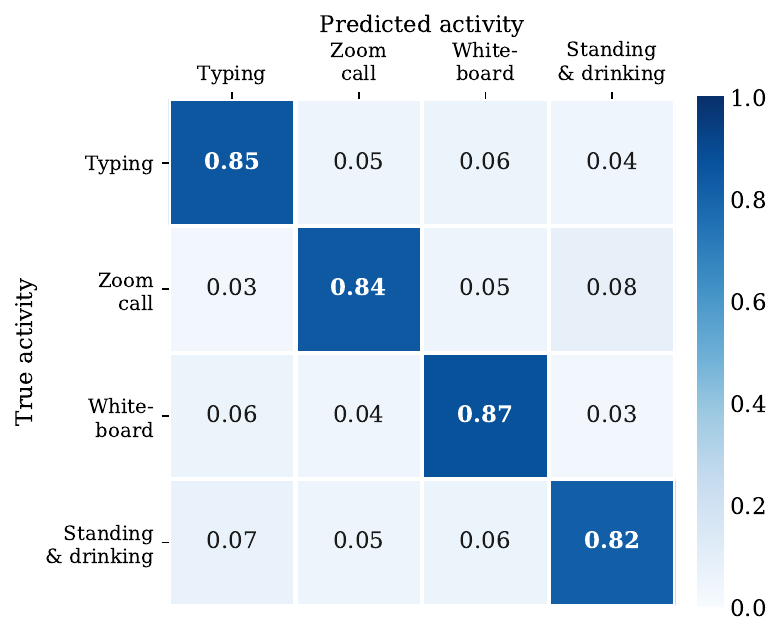}}
\caption{Confusion matrix of our preliminary study.}
\label{fig: exam}
\end{figure}

Since shared sensor data often arrives without labels, we trained a Support Vector Machine~\mbox{\cite{jakkula2006tutorial}} in a self-training fashion: if its soft labels match the true activities, then sharing unlabeled streams already leaks privacy. Figure~\mbox{\ref{fig: exam}} shows the confusion matrix, where the soft labels match the true activities with 82\% to 87\% accuracy. This confirms that activity privacy leakage in multi-sensor environments is real and motivates a protection framework before sharing such streams.

\begin{table}[t!]
\centering
\caption{Characteristics of sensors used in the preliminary study.}
\label{Awair_characteristics}
\begin{tabular}{@{}llrrl@{}}
\toprule
\textbf{Device} & \textbf{Sensor} & \textbf{Range} & \textbf{Error} & \textbf{Unit} \\
\midrule
\multirow{6}{*}{Awair Omni}
 & Humidity     & 0--100      & $\pm$2\%                      & \%   \\
 & Temperature  & $-40$--125  & $\pm$\SI{2}{\degreeCelsius}   & \SI{}{\degreeCelsius} \\
 & $\mathrm{CO}_2$ & 400--5000 & $\pm$75                      & ppm  \\
 & TVOC         & 0--60000    & $\pm$10\%                     & ppb  \\
 & PM2.5        & 0--1000     & $\pm$15                       & \SI{}{\micro\gram\per\cubic\metre} \\
 & SPL          & 48--90      & --                            & dBA  \\
\bottomrule
\end{tabular}
\end{table}

\subsection{Beneficial for IoT Data Training among Users}

For users owning IoT data, the inadequacy of local data often impedes training effective machine learning models, which motivates collaborative learning. Two primary approaches exist. Cloud Computing based Methods~\cite{sunyaev2020cloud,qian2009cloud} aggregate raw user data in the cloud, which improves model performance but exposes raw data to centralized servers. Federated Learning based Methods~\cite{gao2023pfdrl,li2020review} upload only model parameters, yet remain vulnerable to attacks such as model inversion. Stricter safeguards like differential privacy~\cite{husnoo2021differential} further degrade model utility by limiting useful training data.

In practice, the actual sensitivity of data may not align with conventional perceptions. Environmental factors like temperature and CO2 are typically not sensitive, and many activities inferred from them are benign. The real concern arises when innocuous data points are fused to infer sensitive activities, which discourages users from sharing data. This paper therefore proposes a data generation architecture that safeguards privacy-sensitive activities while keeping meaningful activities identifiable, enabling users to confidently share their data.

\begin{figure*}
\centering
      \begin{subfigure}[h]{0.45\linewidth}
      \includegraphics[width=\columnwidth, keepaspectratio= true]{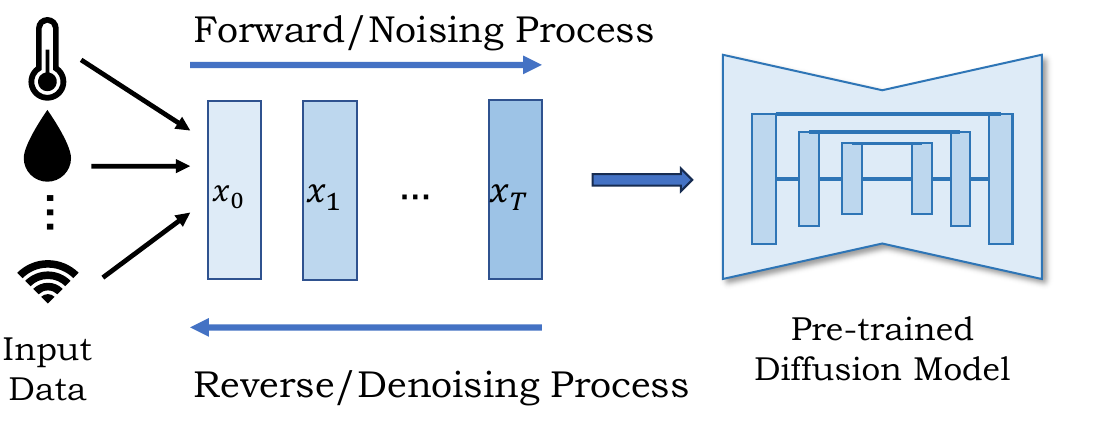}
      \caption{App-conditioned Pre-training Setup}
      \label{fig:set1}
      \end{subfigure}
      \begin{subfigure}[h]{0.45\linewidth}
      \includegraphics[width=\columnwidth, height=3cm, keepaspectratio= true]{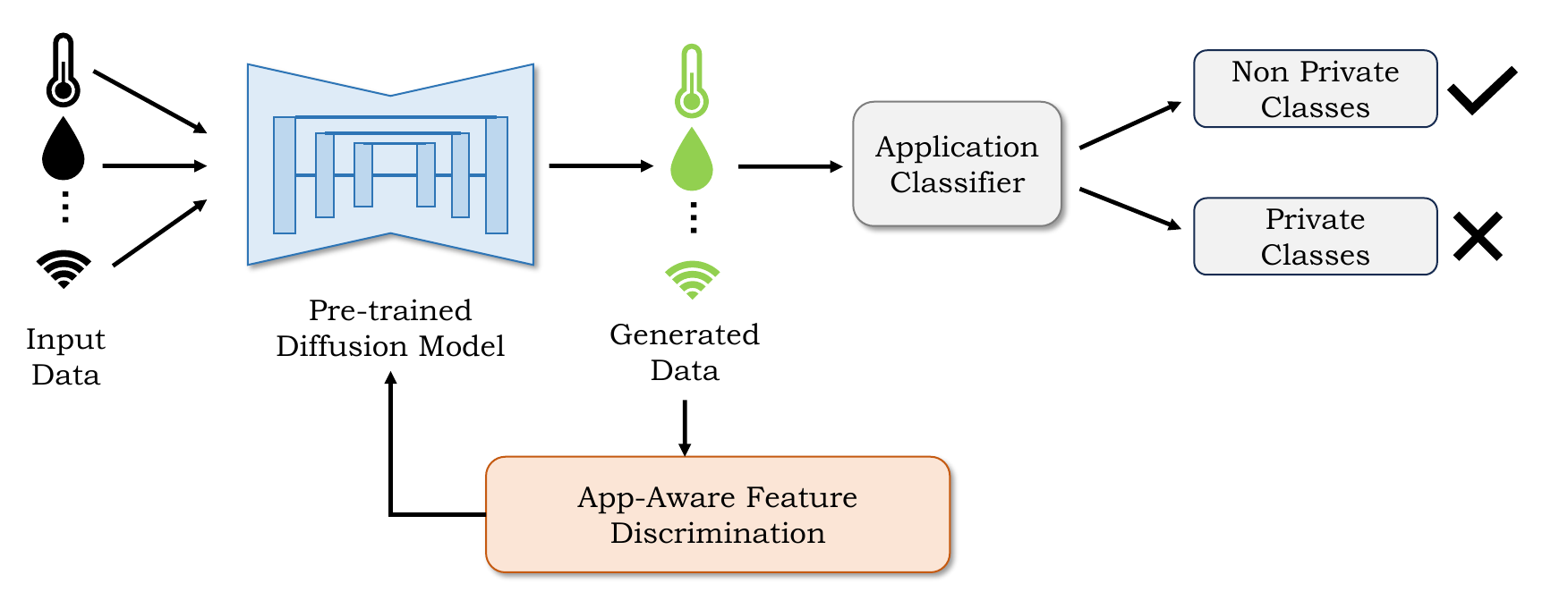}
      \caption{App-aware Fine-tuning Setup}
      \label{fig:set2}
      \end{subfigure}
      
  \caption{Training setups of \ourss. (a) the App-conditioned Pre-training stage; (b) the App-aware Fine-tuning stage.}
  \label{fig:set}
\end{figure*}











\section{Threat Model}
\label{sec:threat}

\subsection{System Model}
\label{sec:system-model}

We consider a sensor-intensive environment, such as a smart office or a smart home, where many sensors continuously record data. A \textit{data owner}, for example an office occupant or a building manager, runs a local IoT hub that collects all of these multi-sensor streams.

The data owner wants to share sensor data with one or more \textit{third-party services}, such as an energy management service or an occupancy analytics provider, so that useful non-private applications keep working. At the same time, the owner does not want to reveal certain activities, which we call \textit{private}, to those services. \ourss~runs locally, on the owner's own device or edge node, before any data leaves the trusted boundary. The raw streams never leave; only the synthetic output of \ourss~is shared. As a result, the owner's own services can still run on raw data with full accuracy, while outside parties only ever see synthetic data.

\subsection{Adversary Model}
\label{sec:adversary}

We treat the adversary as a \textit{third-party service} that receives the synthetic streams produced by \ourss. We consider two levels of attacker. \noindent\textbf{Non-adaptive adversary.} The attacker trains a standard activity classifier on the synthetic data it receives and tries to infer private activities. It does not know that the data was generated by a model. This matches our main evaluation in Section~\ref{sec:rq1}. \noindent\textbf{Adaptive adversary.} The attacker knows the shared data is synthetic and tries to adapt to it, either by retraining its classifier on the synthetic distribution or by using domain adaptation~\cite{ganin2016domain} to close the gap to real data. We study this stronger attacker in Section~\ref{sec:rq5}. We assume the attacker cannot see the raw streams or the internal parameters of \ourss, and cannot infer private activities from timing alone, since the timing of shared streams is controlled by the owner. We measure privacy protection directly through the accuracy of the adversary's classifier on private and non-private activities: lower accuracy on private activities indicates stronger protection, while accuracy on non-private activities close to that of raw data indicates preserved utility.

\section{System Design}



In this section, we illustrate the details of the proposed method.
We draw our motivation from the fact that existing methods mainly focus on how to improve the efficiency of activity prediction, where the privacy issue among these methods are largely neglected, expecting effective techniques to discriminate privacy sensitive data and the non-private sensitive ones. 
Therefore, our method is motivated to leverage the powerful generative ability of diffusion model in discriminating the aforementioned data.
Particularly, we propose a general paradigm to optimize the diffusion model in producing App-aware data, along with two stages, namely App-Conditioned pre-training (ACP) and App-Aware fine-tuning (AAF). Figure~\ref{fig:set} shows the training setup of our method.
Specifically, in Figure~\ref{fig:set1}, ACP pre-trains a diffusion model on multi-sensor data datastreams with task category embedding, so that the model is capable of producing app-conditioned data after this stage. In this process, we aim to build a pre-trained diffusion model for multi-sensor data streams generation and use it as a base model for the AAF step. The steps of ACP follow the regular procedure of diffusion model training. We discuss the detailed steps in the following section~\ref{sec acp}.

Afterward, AAF enables the diffusion model to discriminate privacy-sensitive from non-sensitive data through contrastive learning. Figure~\mbox{\ref{fig:set2}} shows its setup. Given the pre-trained model from ACP, users feed their multi-sensor streams into the diffusion model to generate synthetic data, and an off-the-shelf activity classifier (a simple CNN here, replaceable by any classifier) labels private and non-private classes. We then design an App-aware Feature Discrimination function (AAFD) that uses contrastive learning on the features of real and generated data to update the diffusion model each iteration, so it produces data more related to non-private and less to private applications. Since some iterations yield low-quality generations, we add a data filtering step so that only sufficiently accurate generations reach AAFD. Details are in Section~\mbox{\ref{sec aaf}}.
\subsection{Design Fashion of Pre-training and Fine-tuning}

We adopt the pre-training and fine-tuning fashion for two reasons. \textbf{(1) Accuracy and Efficiency:} pre-trained models learn rich, generalizable representations from large datasets, achieving strong performance with less data and time during fine-tuning. Training end-to-end from scratch is slower and forces the model to handle noise corruption and the contrastive loss at the same time, which hurts accuracy. \textbf{(2) Flexibility:} different users have different privacy standards. For instance, some consider dining private while others do not. Pre-training once and fine-tuning per user lets us quickly generate data tailored to diverse privacy requirements, whereas direct training would restart from scratch for every user demand.

\subsection{App-Conditioned Pre-training}
\label{sec acp}
\begin{figure*}[t!]
\centering
\includegraphics[width=1.8\columnwidth, trim=0 0 0 0]{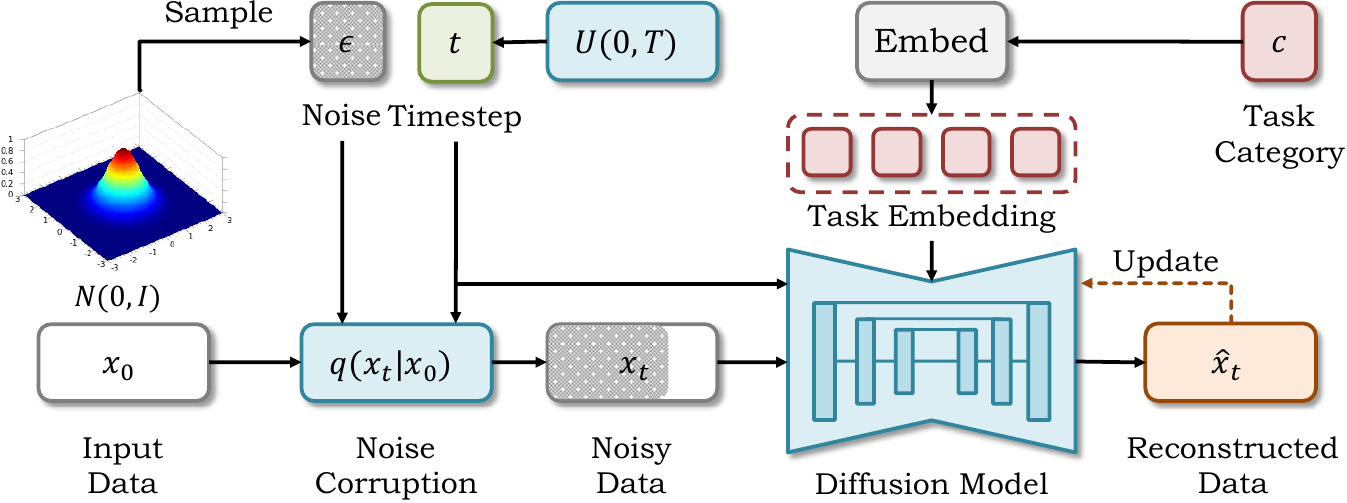}
\caption{
The overall pipeline of the App-Conditioned Pre-training (ACP) stage. The \textbf{output} is a task-conditioned diffusion model that reconstructs clean data \mbox{$\hat{x}_t$} from noisy input, conditioned on the task category embedding.
}
\label{fig: pre-training}
\end{figure*}

Since there is no available pre-trained models to support our motivation, the first step is to train a diffusion model with the multi-sensor data streams, i.e., the ACP stage.
In doing so, ACP trains the diffusion model following the standard procedures, which consist of the \textbf{training} and \textbf{sampling} processes, where Figure~\ref{fig: pre-training} demonstrates the overall pipeline of the ACP stage.

\noindent \paragraph{\textbf{Training}.}
Given an input data $x_0$, we firstly randomly sample 
a noise $\epsilon$ and a timestep $t$ from the random 
Gaussian distribution $N( 0, I )$ and uniform distribution $U( 0, T)$, respectively, where $I$ denotes an identity diagonal matrix and $T$ represents the maximum timestep.
Then, we use the sampled $\epsilon$ and $t$ to corrupt the input clean data $x_0$ into a series of noisy data, termed as $\{x_1, x_2, \dots, x_T\}$, where the process $q(x_t|x_{t-1})$ is formulated by:
\begin{align} \label{eq: reparameterization}
    q(x_t|x_{t-1}) &= N(x_t;\sqrt{1-\beta_t}\cdot x_{t-1}, \beta_t\cdot I), \notag \\
    q(x_{1:T}|x_0) &= \prod^T_{t=1} q(x_t|x_{t-1}),
\end{align}
where the aforementioned equations denote the reparameterization trick in DDPM \cite{ho-etal-2020-ddpm}.
Following Eq. \ref{eq: reparameterization}, we can sample $x_t$ at any timestep $t$ with $\alpha_t = 1 - \beta_t$ and $\bar{\alpha_t} = \prod^t_{i=1} \alpha_i$, where $x_t$ is computed through the following equation:
\begin{align}
    x_t &= q(x_t|x_0), \notag \\
        &= N(x_t; \sqrt{\bar{\alpha_t}} \cdot x_0, (1 - \bar{\alpha_t}) \cdot I), \notag \\
        &= \sqrt{\bar{\alpha_t}} \cdot x_0 + \sqrt{1 - \bar{\alpha}_t} \cdot \epsilon,
\end{align}
where $\sqrt{\bar{\alpha}}_t$ is a blending scalar correlated to the noise schedule of DDPM \cite{ho-etal-2020-ddpm}.

\noindent \paragraph{\textbf{Sampling}.}
The sampling process of diffusion model $p_\theta (x0:T)$ starts from a random Gaussian noise $x_T \sim N(0, I)$, and iteratively de-noises from it into the final clean data $x_0$ conditioned on the task embedding $\mathbf{E}_c$, where the sampling process is formulated as:
\begin{align}
p_\theta(x_{t-1}|x_t) &= N(x_{t-1};\mu_\theta(x_t, t), \Sigma), \notag \\
p_\theta(x_{0:T}) &= p(x_T) \prod^{T}_{t=1} p_\theta (x_{t-1}|x_t),
\end{align}
Specifically, we use DDIM sampler during the sampling process, where the final clean data $x_0$ is generated through:
\begin{align} \label{eq: ddim} 
x_{t-1} 
&=
\sqrt{\Bar{\alpha}_{t-1}} 
\cdot
\frac{x_{t}-\sqrt{1-\Bar{\alpha}_t} \cdot 
\epsilon_\theta(x_t, \mathbf{E}_c, t)}{\sqrt{\Bar{\alpha}_t}}, \notag \\
&+ \sqrt{1-\Bar{\alpha}_{t-1}} 
\cdot
\epsilon_\theta(x_t, \mathbf{E}_c, t),
\end{align}

\subsubsection{Optimization Objective}
In training, the usual variational bound on negative log-likelihood is optimized by:
\begin{align}
    \mathbb{E}[- \log p_\theta(x_0)] \leq \mathbb{E}_q \Big[ -\log \frac{p_\theta (x_{0:T})}{q(x_{1:T}|x_0)} \Big] &= \notag \\
    \mathbb{E}_q \Big[ -\log p(x_T) - \sum \log \frac{p_\theta(x_{t-1}|x_t)}{q(x_t|x_{t-1})} \Big] &= :L,
\end{align}
where $L$ in the equation above is written as:
\begin{align}
L = \mathbb{E}_q
\begin{bmatrix}
    D_{KL} (q(x_t x_0) \Vert p(x_T)) + \\
    \sum_{t \geq 1} D_{KL} (q(x_{t-1} x_t, x_0) \Vert p_\theta (x_{t-1} x_t)) \\
    -\log p_\theta (x_0 x_1)
\end{bmatrix}.
\end{align}

Herein, KL refers to the Kullback-Leibler divergence.

To optimize the diffusion model, we use an embedding layer with a learnable matrix $\mathbf{M}$ to project the task category label $c$ into task embedding $\mathbf{E}_c$, where $\mathbf{E}_c$ then serves as the condition for the diffusion model to de-noise $x_t$. Besides the task category label $c$, we also incorporate statistical features for each task to enrich the information available about each task. The features include the mean, standard deviation, and Z-score, calculated as \((x - \mu) / \sigma\), where \(x\) is an observed value, \(\mu\) represents the mean, and \(\sigma\) denotes the standard deviation.
Once $x_t$ and $\mathbf{E}_c$ are computed, we send $x_t$, $\mathbf{E}_c$, and $t$ into the diffusion model to predict noise, and minimize the Mean-Squared Error (MSE) distance between the predicted and the random Gaussian noise $\epsilon$, where the loss function $L(\theta)$ to optimization the model parameters $\theta$ of the diffusion model is written as:
\begin{equation}
    L(\theta) = \Vert \epsilon - \epsilon_\theta (x_t, t, \mathbf{M}_c) \Vert^2,
\end{equation}
where we follow DDPM in predicting noise $\epsilon_\theta (x_t, t, \mathbf{M}_c)$ instead of the mean, so as to fit the data distribution.
Herein, we conduct the diffusion model following the original architecture of DDPM \cite{ho-etal-2020-ddpm}, which uses an enhanced U-net architecture that consists of an encoder, a decoder, and multiple Transformer \cite{vaswani-etal-2017-transformer} blocks.
Particularly in DDPM, the encoder mainly down-samples the input image into latent representations; the decoder up-samples the latent representations to the original size of image; the Transformer blocks learn the internal correlation between image patches via the self-attention mechanism.
In the setting of multi-sensor environment, the U-net firstly encodes the noisy data $x_t$ into latent representations, and decodes the representations into predicted noise $\epsilon_\theta (x_t, t, \mathbf{M}_c)$, where the Transformer blocks establish the relationship of data at various sensor time.
In this way, we update the model parameters $\theta$ of the diffusion model until convergence, where the optimized diffusion model is capable of producing data according to task categories.

\begin{figure*}[t!]
\centering
\includegraphics[width=1.8\columnwidth, trim=0 0 0 0]{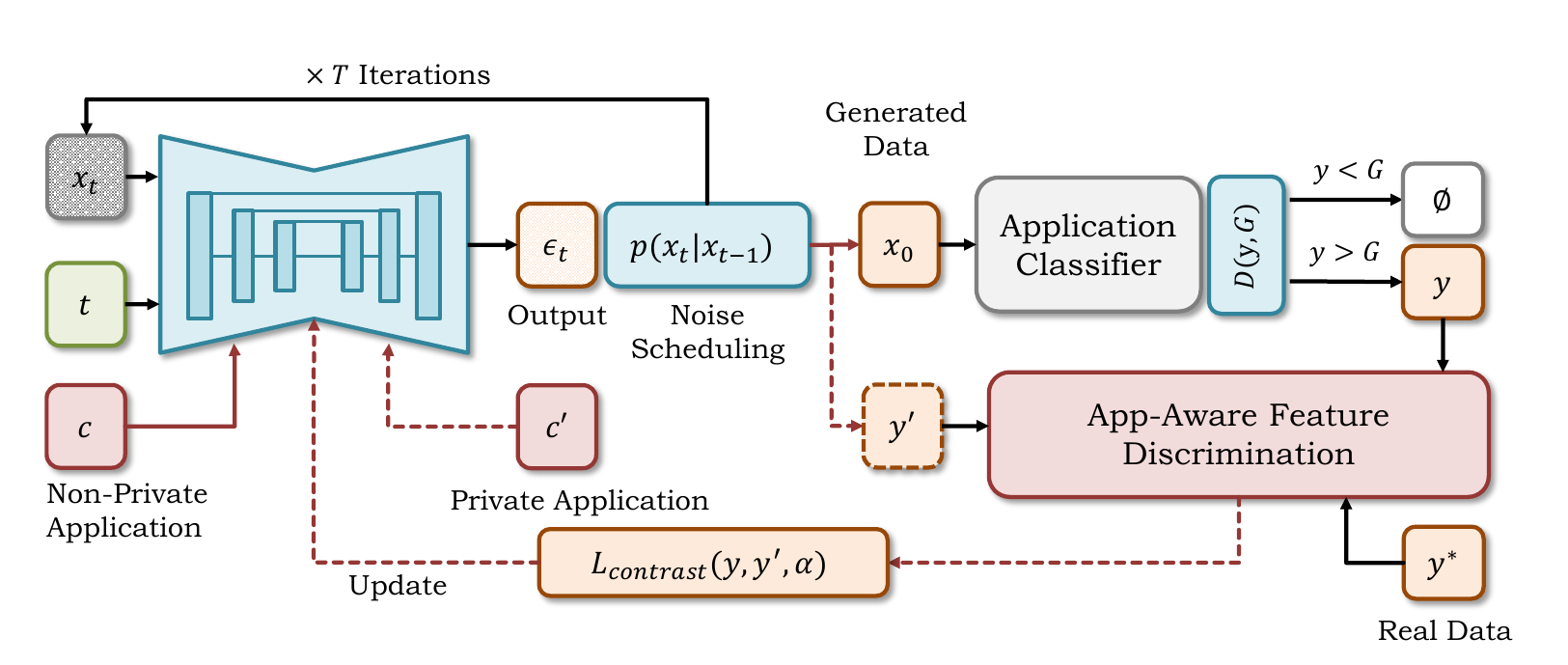}
\caption{
The overall pipeline of the App-Aware Fine-tuning (AAF) stage. The \textbf{output} of this stage is the fine-tuned diffusion model, which generates synthetic data streams (\mbox{$y$}) whose features are pulled toward real non-private data (\mbox{$y^*$}) and pushed away from generated private data (\mbox{$y'$}).
}
\label{fig: fine-tuning}
\end{figure*}

\subsection{App-Aware Fine-tuning}
\label{sec aaf}
With the assistance of the ACP stage, the diffusion model is able to produce realistic data according to various tasks.
Nevertheless, the intrinsic capabilities of the trained model still fails to discriminate privacy sensitive and non-sensitive data.

To improve this, tailored optimization for data discrimination is needed. While Reinforcement Learning with Human Feedback (RLHF)~\cite{bai2022training,casper2023open} aligns model outputs with preferences, RL algorithms, and especially value-based ones like Q-learning~\cite{watkins1992q}, suffer from training instability and high variance, require carefully designed reward functions, demand significant compute, and often generalize poorly to altered tasks or environments. These drawbacks limit their applicability in our resource-constrained, multi-user setting.

Thus, we draw another insight from the advancement of contrastive learning, which digs positive and negative samples in a self-supervised manner, and enables the model to generate outputs that are more similar to the positive samples.
This technique has illustrated promising performance in a wide series of down-stream tasks in the computer vision community, and shows superior efficiency compared to RLHF without the extra requirements of human feedback. Figure~\ref{fig: fine-tuning} shows the overall pipeline of our AAF stage.
To enhance the diffusion model with contrastive learning, we need to consider from two perspectives: (1) the rules of selecting positive and negative samples; (2) the loss function of contrastive learning.
In the following texts, we present the details of both aforementioned perspectives.

\paragraph{\textbf{Rules to Select Positive and Negative Samples}.}
Our goal of contrastive learning is to enable the diffusion model to discriminate privacy sensitive and non-sensitive data.
Therefore, the rules to select positive and negative samples are straightforward, where we choose the real non-sensitive data as positive sample, and treat generated data with respect to privacy sensitive task categories.
In details, given the privacy sensitive task category $c'$ and non-sensitive one $c$, we follow the standard process of Eq. \ref{eq: ddim} and generate the corresponding data with the pre-trained diffusion model, termed as $x_0$ and $y'$, respectively.
Then, we randomly sample a real non-sensitive data $y^*$ from the multi-sensor datas streams $\mathcal{D} = \{ y^*_1, y^*_2, \dots, y^*_{N_D} \}$ with the total length of dataset termed as $N_D$.
In this way, we leverage $y^*$ and $y'$ as the positive and negative samples, respectively, where $x_0$ and $y'$ are then used in later processes of the AAF stage.

\paragraph{\textbf{Data Filtering}.}
Before we compute the loss function with contrastive learning, it is vital to ensure the data quality of $x_0$.
Once we optimize the diffusion model with ill-presenting quality of $x_0$, there are possibilities that $x_0$ has closer distance to $y'$ than $y^*$, where might lead to inferior optimization during fine-tuning.
To solve this problem, we leverage an off-the-shelf activity classifier $\mathcal{E}_\phi$ with its model parameters as $\phi$.
We observe that low-quality data would cause $\mathcal{E}_\phi$ to produce inaccurate results, where we set a threshold $G$ to filter the produced low-quality data, where the data filtering process $D(x_0, G)$ is formulated as:
\begin{equation}
\label{eq: data-filter}
y = D(x_0, G) = \left\{
\begin{aligned}
    x_0&, \quad \quad \mathcal{E}_\phi (x_0) > G \\
    \emptyset&, \quad \quad \mathcal{E}_\phi (x_0) \leq G
\end{aligned}
\right.
\end{equation}
Afterward, we use the filtered data $y$ to compute the loss function of contrastive learning.
If the data filtering process outputs $\emptyset$, we do not compute the loss function of contrastive learning in this training iteration. In our experiment, we use a simple convolutional neural network (CNN) for application classification, which contains four convolutional layers followed by a ReLU activation function with 2 fully connected layers. 


\textbf{Choice of $G$.} The threshold $G$ controls the trade-off between filtering strictness and the amount of training signal that survives. A low $G$ lets low-quality generations through and can corrupt the contrastive objective, while a high $G$ rejects too many samples and starves the fine-tuning step. We set $G$ from the classifier confidence distribution on a held-out validation set, choosing a value near the median confidence of correctly classified real samples, which in our data falls in the range $[0.6, 0.8]$. All main experiments use $G = 0.7$. Section~\ref{sec:rq5} (Figure~\ref{fig:threshold}) sweeps $G$ from $0.5$ to $0.9$ and shows that performance is stable within the $[0.6, 0.8]$ band.

\textbf{Decoupling filtering, training, and evaluation classifiers.} To avoid a circular evaluation, the off-the-shelf classifier $\mathcal{E}_\phi$ used for filtering and feature extraction during fine-tuning is independent of the classifiers used to measure privacy protection in Section~\ref{sec:evaluation}. The evaluation attacker always trains its own classifier from scratch on the synthetic data. Section~\ref{sec:rq1} further shows that the protection holds across three different classifier architectures (CNN, Transformer, and Random Forest), so the effect is not tied to one specific model.

\begin{figure}[t!]
\centering
\includegraphics[width=\linewidth, trim=0 0 0 0]{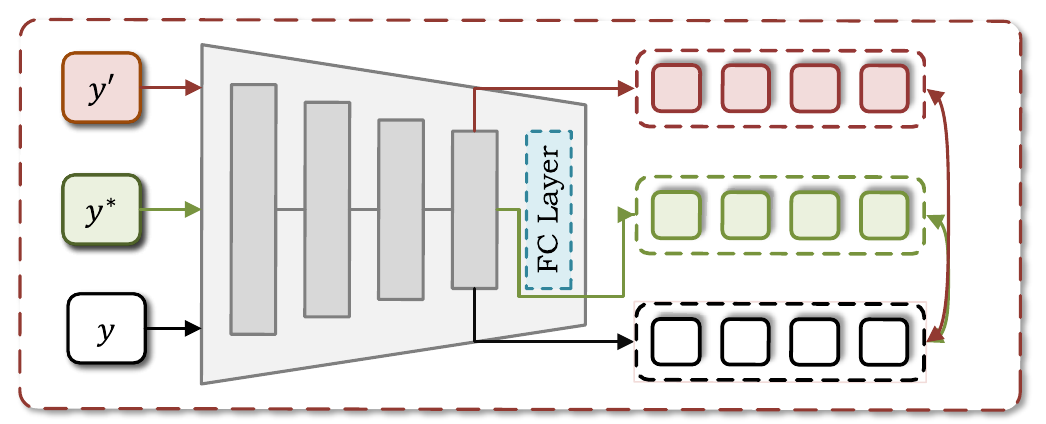}
\caption{
The overall pipeline of the App-Aware Feature Discrimination (AAFD).
Herein, the model in gray represents the off-the-shelf activity classifier; ``FC'' denotes the fully-connected layer, which is normally a single linear projection layer to project the final latent representations into a probability distribution.
}
\label{fig: feature-discrimination}
\end{figure}

\subsubsection{App-Aware Feature Discrimination}
Once $y$, $y'$, and $y^*$ are ready, the final step is to optimize the diffusion model with the contrastive learning, i.e., the App-Aware Feature Discrimination (AAFD) process.
Figure~\ref{fig: feature-discrimination} shows the overall pipeline of AAFD, where the model in gray represents the off-the-shelf activity classifier.
To perform AAFD, we use the activity classifier $\mathcal{E}_\phi$ as a feature extractor, where we leverage the latent representations before the final Fully-Connected (FC) layer as extracted features, and then conduct the contrastive learning by comparing various features of different data, encouraging the diffusion model to generate outputs that are more similar to $y^*$ rather than $y'$.
Specifically, we firstly input $y$, $y'$, and $y^*$ into $\mathcal{E}_\phi$, and obtain their corresponding features, termed as $\mathbf{F}$, $\mathbf{F}'$, and $\mathbf{F}^*$, respectively.
Then, we compute the loss function of contrastive learning (termed $L_{contrast} (\theta)$) following the standard form of triple loss, which is written as the following equation:
\begin{equation} \label{eq: contrastive-loss}
L_{contrast} (\theta) =
\big[ \Vert \mathbf{F} - \mathbf{F}^* \Vert^2 - \Vert \mathbf{F} - \mathbf{F'} \Vert^2 + \alpha \big],
\end{equation}
where $\alpha$ represents a hyper-parameter that determines the margin value of the triplet loss.
In training, we optimize the diffusion model with the loss function in Eq. \ref{eq: contrastive-loss} until convergence.
For further analyses of this stage, the contrastive learning encourages the diffusion model to produce data $y$ that have closer feature distance to the real data $y^*$ rather than $y'$, and allow the activity classifier $\mathcal{E}_\phi$ to output more accurate results with the corresponding feature inputs.
In this way, we are capable of allowing the diffusion model to producing privacy non-sensitive data, meanwhile maintaining the sample quality in its generated data.

\section{Evaluation}
\label{sec:evaluation}
To evaluate the performance of \ourss~on generality, practicability and usefulness, we experiment our method on real-world smart home and smart office multi-sensor datasets on the following research questions:

\begin{itemize}
    \item \textbf{RQ1}: 
    How does \ourss~perform on synthetic data generation? Can the generated data streams from \ourss~detect non-private applications and conceal private ones? How does it compare against both generative baselines and simple replay and perturbation baselines, and is the protection consistent across different classifier architectures?
    \item \textbf{RQ2}: How does \ourss~perform if the users switch the selection of the private/non-private applications? Will it influence the overall performance? This is a practical test to see if \ourss~is robust enough to handle different users.
    \item \textbf{RQ3}: How do the generated data streams perform on their original purposes? (e.g., $\mathrm{CO}_2$ prediction, temperature prediction, etc)  
    \item \textbf{RQ4}: How does \ourss~perform if there are missing application labels? 
    \item \textbf{RQ5}: How robust is \ourss~against an adaptive adversary that knows the data is synthetic and retrains its classifier on the synthetic distribution? And how sensitive is \ourss~to the filtering threshold \mbox{$G$}?
\end{itemize}

\subsection{Applications and Datasets}

We validate \ourss~on three real-world datasets spanning three sensing scenarios, to show it is not tied to one sensor type or environment.

\textbf{(1) Smart Home (CASAS)}~\cite{alam2023fedaiot}: the CASAS dataset\footnote{https://casas.wsu.edu/datasets/} records daily living activities through sequences of sensor states across apartments equipped with motion, temperature, light, water, burner, and door sensors. We keep the pre-processing of~\cite{alam2023fedaiot} and use five datasets (Milan, Kyoto1--4) chosen for consistent sensor representation. We streamline the original categories into 11 home-activity categories such as sleep, eat, leave home, enter home, and bath. Each entry is a categorical time series of length 2{,}000. We label \textit{bath} and \textit{leave home} as private. 

\textbf{(2) Smart Office} (self-collected): the setup from Section~\ref{sec:motivation} (Figure~\ref{fig: uva}), with AWAIR Omni\textsuperscript{\textcopyright} sensors in two rooms, five participants, and four activities (typing, Zoom calls, whiteboard writing, standing while drinking), where \textit{Zoom calls} are private.

\textbf{(3) In-the-Wild (ExtraSensory)}~\cite{vaizman2017recognizing}: collected from 60 users with their own smartphones and smartwatches over about a week each, totaling more than 300{,}000 labeled minutes. Each minute carries phone and watch motion, location, audio, and ambient measurements with self-reported context labels such as walking, sitting, sleeping, cooking, and shower. Being in the wild, it is noisier and far less balanced than the indoor datasets, and naturally contains missing sensors and labels, which is exactly the situation studied in RQ4. We label \textit{shower} and \textit{sleeping} as private. Because it is in the wild, its raw accuracy is lower and variance higher than the indoor datasets.


\subsection{Experimental Setup and Evaluation Metrics}
We implemented our experiments using one NVIDIA RTX 3090 GPU. For the diffusion model settings, the step size of the forward diffusion process is controlled by a variance schedule $\beta_t \in (0.0001, 0.05)$, where $t$ ranges from 1 to $T$. The maximum diffusion step is set to be 100. The batch size was 128, and training spanned 200 epochs with early stopping using a patience of 30 epochs to counter overfitting. 

For the CASAS datasets, since we are under the pre-training and fine-tuning setup, we combine these datasets together for training in a leave-one-out fashion. For example, If we assume "Kyoto4" is a new user, we use “Milan”, “Kyoto1”,“Kyoto2” and “Kyoto3” to perform the pre-training step, then only use "Kyoto4" for fine-tuning and testing. We divided the leave-one-out dataset into fine-tuning training and testing sets using an 80-20 split.  

For our self-collected smart office dataset, since we collected the multi-sensor data streams on 2 office rooms, we chose to pre-train on one room, and fine-tuning on other room, and we also divided the fine-tuning room datasets into fine-tuning training and testing sets using an 80-20 split.  

For comparison, since \ourss~is a synthetic data generation framework, we compare against methods of the same functionality. Most GAN-based models~\cite{hu2023bsdgan,wang2018sensorygans,yang2023ts} suffer from limited diversity, mode collapse, and unstable convergence, so we select GAN baselines tailored to our setting, plus two simple non-learning baselines:

\begin{itemize}
    \item \textbf{TimeGAN~\mbox{\cite{yoon2019time}}:} A GAN known for stable training on time-series data. We replace only the diffusion model with TimeGAN under our pre-training and fine-tuning setup.
    \item \textbf{Conditional GAN~\mbox{\cite{mirza2014conditional}}:} Generates data conditioned on labels. We condition it to conceal private activities by assigning the discriminator a lower score on private activities and a higher score on non-private ones. Since the condition is embedded in training, we use direct training.
    \item \textbf{Occupancy-GAN~\mbox{\cite{wu2021smart}}:} Uses two application classifiers (\mbox{$C_1$} and \mbox{$C_2$}) that feed two losses into the generator, optimizing \mbox{$C_1$} while suppressing \mbox{$C_2$}. Since the condition is embedded in training, we use direct training.
    \item \textbf{Replay (simple baseline):} Replaces each private segment with a randomly sampled real non-private segment, and leaves non-private segments untouched. It requires no learning.
    \item \textbf{Noise Perturbation (simple baseline):} Adds calibrated Gaussian noise to all streams, tuned so that private accuracy matches \ourss, to test whether uniform perturbation can reach the same privacy-utility trade-off.
\end{itemize}


The private/non-private labels above follow prior privacy work (presence, hygiene, and meeting content are treated as sensitive~\cite{wu2021smart}). Importantly, \ourss~treats this designation as a user-defined input decoupled from the framework, and RQ2 evaluates what happens when it is switched.

For evaluation metrics, we use accuracy for the classification tasks, repeating each experiment 5 times and averaging. We report accuracy on both private and non-private activities: lower private accuracy indicates stronger protection, while non-private accuracy close to raw data indicates preserved utility.

\begin{table*}
\centering
\caption{Performance Metrics Comparison across all datasets. For each dataset, the \emph{Non-private} row reports accuracy on activities meant to remain detectable, while the private rows (Bath, Leave home, Zoom call, Shower, Sleeping) report accuracy on activities the data owner wants concealed. Lower private accuracy together with higher non-private accuracy indicates better privacy protection. Values are accuracy \mbox{$\pm$} standard deviation over 5 runs.}
\scalebox{0.85}{
\begin{tabular}{llccccc}
\toprule
\textbf{Dataset} & \textbf{Activity Acc} & \textbf{TimeGAN~\cite{yoon2019time}} & \textbf{Conditional GAN~\cite{mirza2014conditional}} & \textbf{Occupancy-GAN~\cite{wu2021smart}} & \textbf{PrivateHub} & \textbf{Raw data} \\
\midrule
\multirow{3}{*}{Milan} 
 & Non-private & 0.774 $\pm$ 0.052 & 0.653 $\pm$ 0.093 & 0.692 $\pm$ 0.032 & 0.840 $\pm$ 0.067 & 0.84 \\
 & Bath & 0.425 $\pm$ 0.045 & 0.498 $\pm$ 0.025 & 0.512 $\pm$ 0.020 & 0.380 $\pm$ 0.019  & 0.78\\
 & Leave home & 0.471 $\pm$ 0.021 & 0.495 $\pm$ 0.035 & 0.510 $\pm$ 0.022 & 0.393 $\pm$ 0.023 &  0.83\\
\midrule
\multirow{3}{*}{Kyoto1} 
 & Non-private & 0.686 $\pm$ 0.038 & 0.622 $\pm$ 0.050 & 0.539 $\pm$ 0.046 & 0.778 $\pm$ 0.064  & 0.80\\
 & Bath & 0.421 $\pm$ 0.043 & 0.494 $\pm$ 0.034 & 0.502 $\pm$ 0.022 & 0.365 $\pm$ 0.016 & 0.72\\
 & Leave home & 0.441 $\pm$ 0.037 & 0.492 $\pm$ 0.019 & 0.500 $\pm$ 0.015 & 0.381 $\pm$ 0.023 & 0.81\\
\midrule
\multirow{3}{*}{Kyoto2} 
 & Non-private & 0.760 $\pm$ 0.027 & 0.635 $\pm$ 0.026 & 0.663 $\pm$ 0.019 & 0.842 $\pm$ 0.023  & 0.83\\
 & Bath & 0.415 $\pm$ 0.039 & 0.497 $\pm$ 0.029 & 0.510 $\pm$ 0.024 &  0.286 $\pm$ 0.068  & 0.74\\
 & Leave home & 0.438 $\pm$ 0.033 & 0.490 $\pm$ 0.021 & 0.495 $\pm$ 0.016  & 0.392 $\pm$ 0.033  & 0.87\\
\midrule
\multirow{3}{*}{Kyoto3} 
 & Non-private & 0.688 $\pm$ 0.041 & 0.612 $\pm$ 0.025 & 0.533 $\pm$ 0.039 & 0.725 $\pm$ 0.025   & 0.84\\
 & Bath & 0.410 $\pm$ 0.045 & 0.496 $\pm$ 0.031 & 0.507 $\pm$ 0.025 & 0.239 $\pm$ 0.097  & 0.70\\
 & Leave home & 0.432 $\pm$ 0.030 & 0.487 $\pm$ 0.023 & 0.499 $\pm$ 0.017 & 0.301 $\pm$ 0.014  & 0.78\\
\midrule
\multirow{3}{*}{Kyoto4} 
 & Non-private & 0.767 $\pm$ 0.043 & 0.627 $\pm$ 0.041 & 0.683 $\pm$ 0.034 & 0.841 $\pm$ 0.030 & 0.83\\
 & Bath & 0.418 $\pm$ 0.042 & 0.491 $\pm$ 0.028 & 0.508 $\pm$ 0.020 & 0.376 $\pm$ 0.055 & 0.76\\
 & Leave home & 0.435 $\pm$ 0.036 & 0.484 $\pm$ 0.018 & 0.503 $\pm$ 0.014& 0.345 $\pm$ 0.026 & 0.80\\
\midrule
\multirow{2}{*}{Smart office} 
 & Non-private & 0.812 $\pm$ 0.022 & 0.721 $\pm$ 0.026 & 0.821 $\pm$ 0.031 & 0.953 $\pm$ 0.027 & 0.95 \\
 & Zoom call & 0.421 $\pm$ 0.029 & 0.545 $\pm$ 0.034 & 0.428 $\pm$ 0.037 & 0.396 $\pm$ 0.046 &   0.93\\
\midrule
 \multirow{3}{*}{ExtraSensory} 
 & Non-private & 0.712 $\pm$ 0.054 & 0.598 $\pm$ 0.061 & 0.643 $\pm$ 0.049 & 0.781 $\pm$ 0.052 & 0.79 \\
 & Shower & 0.428 $\pm$ 0.051 & 0.495 $\pm$ 0.034 & 0.509 $\pm$ 0.031 & 0.361 $\pm$ 0.038 & 0.76 \\
 & Sleeping & 0.441 $\pm$ 0.045 & 0.491 $\pm$ 0.031 & 0.503 $\pm$ 0.027 & 0.382 $\pm$ 0.040 & 0.78 \\
\bottomrule
\end{tabular}
}
\label{milan}
\end{table*}

\subsection{Results}

\subsubsection{RQ1: Performance of synthetic data generation}
\label{sec:rq1}

In this section, we compare how \ourss~performs against other methods of concealing private activities. Table~\mbox{\ref{milan}} shows that \ourss~matches raw-data accuracy on non-private activities while pushing private accuracy below 40\%, about 40 to 50\% lower than raw. TimeGAN does better than Conditional GAN and Occupancy-GAN on both sides, but its non-private accuracy is well below \ourss~and its private accuracy is higher, though all methods lower private accuracy relative to raw. The reason is that the pre-training and fine-tuning setup smooths training so the non-private features stay close to the raw distribution, while our AAFD better separates private from non-private activities and drives private accuracy much lower.

On the other hand, \ourss~outperforms TimeGAN. Although we only swap the diffusion model for TimeGAN, the GAN architecture is less suited to this task. Diffusion models generate higher-quality samples through incremental denoising and are more stable to train because of their Markov-chain probabilistic foundations, which avoids the convergence difficulties common to GANs. This produces synthetic data that more closely matches the real distribution.

\paragraph{\textbf{Comparison with simple baselines.}}

\begin{figure}[t!]
\centering
\includegraphics[width=\linewidth, trim=0 0 0 0]{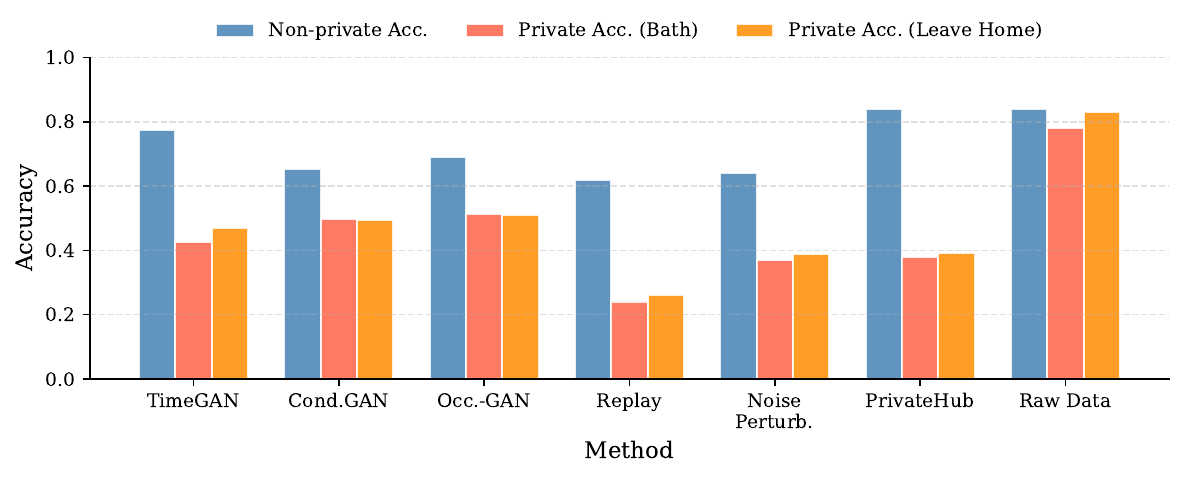}
\caption{Comparison against simple baselines (Replay and Noise Perturbation) on the Milan dataset. The simple baselines either fail to preserve non-private utility (Replay) or require indiscriminate degradation (Noise). \ourss~preserves non-private accuracy while concealing private activities.}
\label{fig:simple-baselines}
\end{figure}

A central question is whether a generative framework is really needed, or whether a simpler strategy such as data replay or noise perturbation would do. Figure~\ref{fig:simple-baselines} compares both simple baselines against \ourss~on the Milan dataset. The Replay baseline does hide private activities well, dropping private accuracy to near random in the 0.24 to 0.26 range, but it pays a steep price. Substituting private segments with unrelated non-private segments breaks the natural timing and the correlations across sensors, so non-private accuracy falls to 0.618, a 26\% relative drop from raw. This violates the first requirement, accurate identification of non-private applications. The Noise Perturbation baseline, when tuned to match the private accuracy of \ourss~at about 0.38, lowers non-private accuracy to 0.641 and, as RQ3 shows, also harms downstream CO$_2$ and temperature prediction, because the noise is applied to every stream at once. In contrast, \ourss~keeps non-private accuracy at 0.840, matching raw data, while reaching comparable private concealment. The key challenge is therefore not to reduce inference accuracy, but to reduce it selectively while keeping the data useful, which is exactly what the simple baselines cannot do.

\paragraph{\textbf{Generalizability across classifiers.}}

\begin{figure}[t!]
\centering
\includegraphics[width=0.82\linewidth, trim=0 0 0 0]{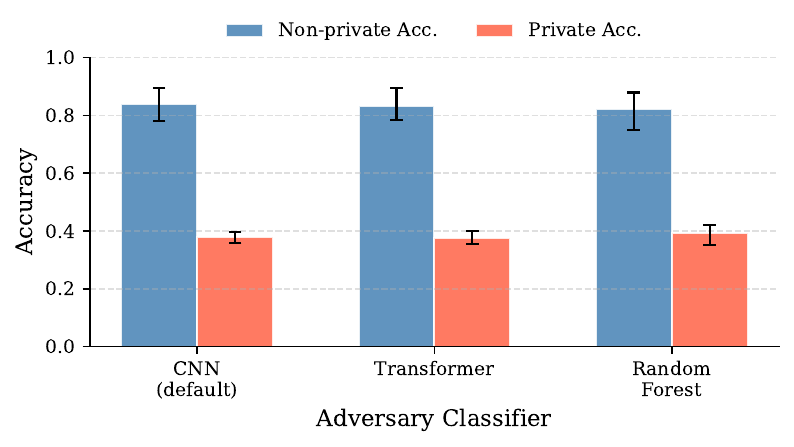}
\caption{Cross-classifier evaluation on Milan. The protection, shown by low private accuracy and high non-private accuracy, holds across three independent attacker classifiers.}
\label{fig:cross-classifier}
\end{figure}

A fair concern is evaluation circularity, where the same classifier used for filtering and feature extraction is also used to measure privacy. To address this, we re-measure the protection using three independent attacker classifiers (CNN, Transformer, and Random Forest), each trained from scratch. As Figure~\ref{fig:cross-classifier} shows, the protection holds across all three, with private accuracy in the 0.37 to 0.39 range and non-private accuracy in the 0.82 to 0.84 range, confirming it comes from the structure of the generated data rather than one particular classifier.

\subsubsection{RQ2: Performance of synthetic data generation with switched private and non-private activities}
\label{sec:rq2}
We compare the robustness of \ourss~by switching the private and non-private activity labels. By default, we use "bath" and "leave home" as private activities, and the rest we label as non-private. However, different users may have different privacy activities. In this experiment, we switch the labels 3 times and we demonstrate the performance on "Milan" and "Kyoto1" datasets. Private 1 we labeled "cook" and "eat" as private, Private 2 we labeled "personal hygiene" and "eat" as private, Private 3 we kept the "bath" and "leave home" to see the difference.

\begin{figure}[t!]
\centering
\includegraphics[width=\linewidth, trim=0 0 0 0]{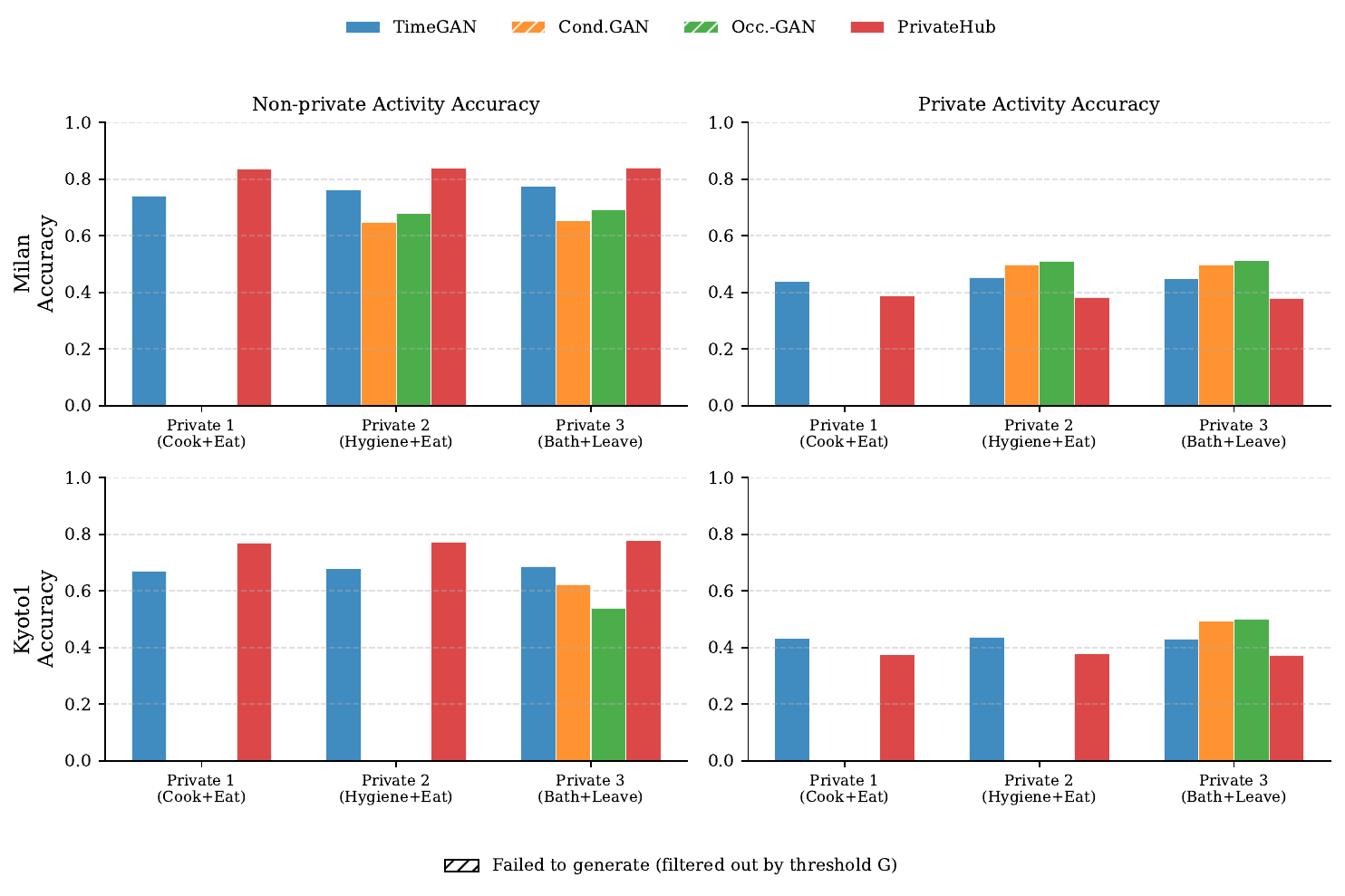}
\caption{Performance comparison when switching private and non-private designations. Left column: non-private accuracy (higher is better). Right column: private accuracy (lower is better). Top row: Milan; bottom row: Kyoto1. Hatched bars mark settings where a method failed to produce enough samples that clear the filtering threshold \mbox{$G$}.}
\label{fig:sw}
\end{figure}

Figure~\ref{fig:sw} shows the result of the private and non-private switch, which follows the same overall trend as RQ1. For Milan, the GAN baselines fail to generate qualified data under the Private~1 setting, and for Kyoto1 they fail under both Private~1 and Private~2, while \ourss~succeeds under all three settings. Here, ``fail to generate'' has a precise meaning tied to our data filtering step. Recall from Section~\ref{sec aaf} that a generated sample is only kept for training if the activity classifier assigns it a confidence above the threshold $G$ (Eq.~\ref{eq: data-filter}). Under these settings the GAN baselines suffer from mode collapse on the smaller and more imbalanced label sets, so almost none of their generated samples clear $G$, and there are not enough qualified samples to complete the fine-tuning step. We therefore mark the corresponding bars as failed. This is most visible on Kyoto1, which has very few activity occurrences, so the effect is strongest there. \ourss~does not run into this problem, because the diffusion model produces more diverse samples and a healthy fraction of them clear $G$ under every setting. In addition, instead of relying on conditional embedding, our contrastive AAFD loss pulls generated samples toward non-private real data and away from generated private data, which helps the model produce more non-private and less private content.

\subsubsection{RQ3: Performance of synthetic data generation with original purpose}

\begin{figure}[t!]
\centering
\includegraphics[width=\columnwidth, trim=0 0 0 0]{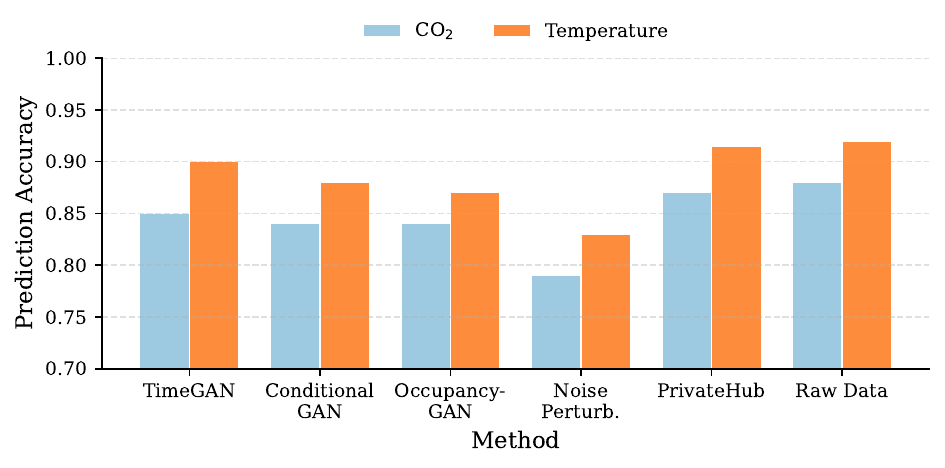}
\caption{Downstream prediction on $\mathrm{CO}_2$ and temperature: the Noise Perturbation baseline degrades utility the most, while \ourss~stays closest to raw data.}
\label{cco2}
\end{figure}

In this section, we check whether the generated data damages the original purpose of each stream. For our collected data, the AWAIR Omni\mbox{\textsuperscript{\textcopyright}} sensors monitor indoor environmental quality and support tasks such as \mbox{$\mathrm{CO}_2$} and temperature prediction, which the generation could harm. Figure~\mbox{\ref{cco2}} shows the result using a simple LSTM~\mbox{\cite{graves2012long}}. All methods give reasonable predictions, and \ourss~is closest to raw data.


The Noise Perturbation baseline performs worst on these downstream tasks. Because it must add substantial noise to every stream to conceal private activities, it lowers $\mathrm{CO}_2$ prediction accuracy to 0.79 and temperature prediction to 0.83, both well below \ourss. This supports the finding from RQ1 that indiscriminate perturbation cannot keep the sensor data meaningful, which is our third requirement, while the targeted generation of \ourss~can.



\begin{figure}[t!]
\centering
\includegraphics[width=\columnwidth, trim=0 0 0 0]{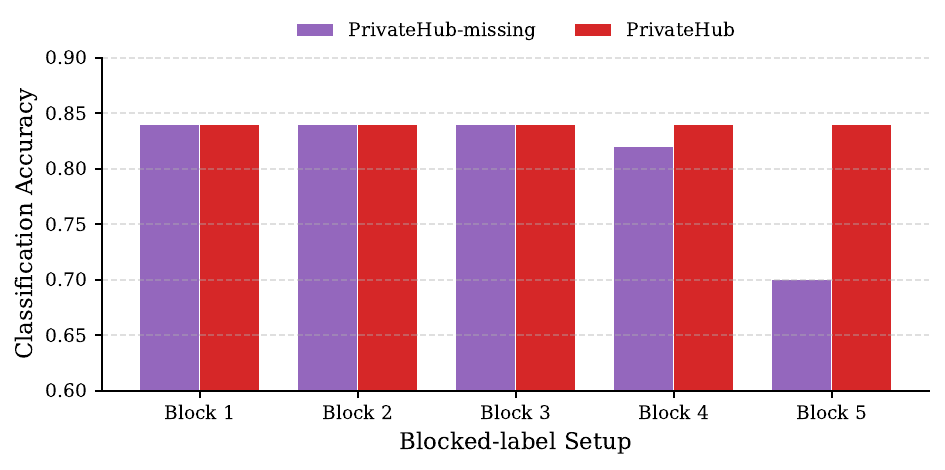}
\caption{Missing application labels: accuracy is preserved in most setups, and only Block~5, a label highly correlated with private activities, shows a notable drop.}
\label{block}
\end{figure}

\subsubsection{RQ4: Performance of \ourss~on missing application labels}

In reality, it's very likely that when the user sends the data streams, there will be missing activity labels. However, it's crucial to understand whether the generated data from \ourss~with partial activity labels can still be classified. In this experiment (Figure~\ref{block}),  we block 2 activity labels and train \ourss~with all other labels then test if we can still identify the blocking 2 applications. We tried 5 different setups and compared \ourss~with the missing label training version. We observe that for most (60\%) of the setups, we have no accuracy lost. For 20\% of the setups, we have around 2\% accuracy drop. For the last setup, we have around a 14.2\% accuracy drop. The reason is probably that the blocking label is highly related to private applications.

The ExtraSensory dataset offers a more realistic version of this test, since it is collected in the wild and already contains missing labels and missing sensors rather than blocks we remove by hand. When we train \ourss~on its partially labeled streams, we observe the same overall behavior as in the controlled block experiments: most activities are still identified with little loss, and the larger drops occur only for labels that are closely tied to private activities. This indicates that the robustness of \ourss~to missing labels is not an artifact of how we construct the controlled blocks, but holds under naturally incomplete data as well.

\subsubsection{RQ5: Robustness to Adaptive Adversaries and Threshold Sensitivity}
\label{sec:rq5}

\paragraph{\textbf{Adaptive adversary.}}

\begin{figure}[t!]
\centering
\includegraphics[width=0.95\linewidth, trim=0 0 0 0]{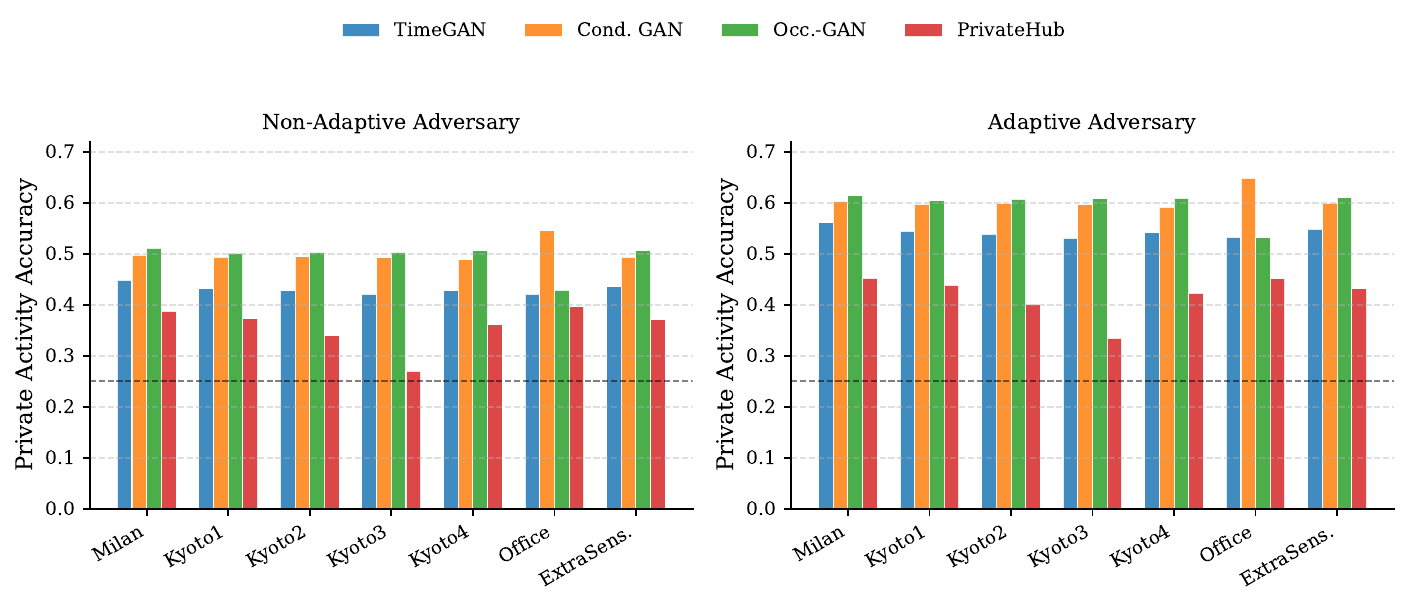}
\caption{Private-activity accuracy under a non-adaptive attacker (left) and an adaptive attacker that retrains on the synthetic distribution (right). Lower is better; the dashed line marks the random-guess baseline. \ourss~recovers the least under adaptation, which indicates feature-space rather than surface-level protection.}
\label{fig:adaptive}
\end{figure}

The evaluations above assume a non-adaptive adversary that treats the synthetic data as if it were real. A stronger and more realistic threat is an adaptive adversary (Section~\ref{sec:adversary}) that knows the data is synthetic and retrains its classifier directly on the synthetic distribution to try to recover private activities. Figure~\ref{fig:adaptive} compares the private-activity accuracy of all methods under the non-adaptive attacker on the left and the adaptive attacker on the right.

Under the adaptive attacker, all methods recover some private-activity accuracy, but by very different amounts. For the GAN baselines, private accuracy rises sharply. TimeGAN on Milan goes from 0.45 to 0.56, an 11 point jump, because its concealment is mostly a surface change to the data that the attacker can re-learn once it adapts to the synthetic distribution. \ourss~is far more resilient. Its private accuracy on Milan rises only from 0.39 to 0.45, a 6 point jump, and stays well below both the GAN baselines and the raw-data accuracy. The reason is that AAFD does not just perturb the surface of the data. It restructures the data in the feature space of an activity classifier, pulling the representations of private activities toward those of non-private ones. Since the two are entangled rather than merely shifted, retraining on the synthetic distribution cannot easily pull them apart. \ourss~therefore offers a structurally more robust form of protection than methods that only change the output distribution.

\paragraph{\textbf{Sensitivity to threshold $G$.}}

\begin{figure}[t!]
\centering
\includegraphics[width=0.9\linewidth, trim=0 0 0 0]{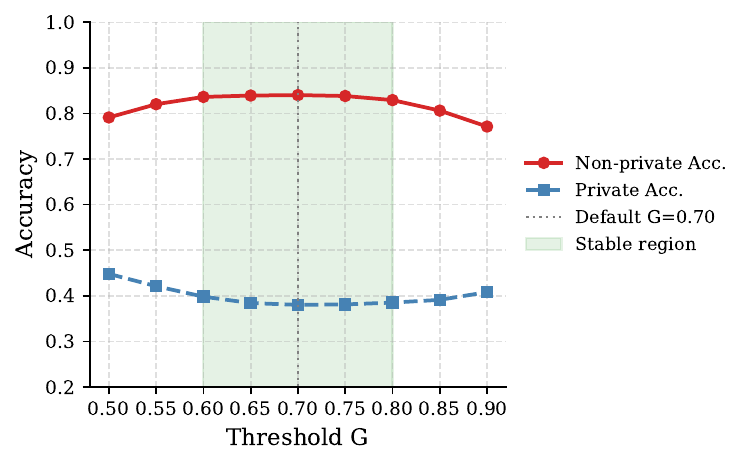}
\caption{Sensitivity to the filtering threshold \mbox{$G$} on Milan. Performance is stable within the shaded \mbox{$[0.6,0.8]$} region, and the default \mbox{$G=0.7$} is marked.}
\label{fig:threshold}
\end{figure}

We study how the data-filtering threshold $G$ affects the privacy-utility trade-off by sweeping $G$ from 0.5 to 0.9 on Milan (Figure~\ref{fig:threshold}). When $G$ is too low (around 0.5), low-quality generations leak into the contrastive objective, which slightly raises private accuracy and lowers non-private accuracy. When $G$ is too high (around 0.9), too many samples are filtered out, which starves the fine-tuning signal and again lowers non-private accuracy. Within the $[0.6, 0.8]$ band both metrics are stable, supporting the default $G=0.7$, so the method is not brittle to the exact value of $G$ as long as it lies in a reasonable range set by the validation confidence distribution.

\subsubsection{Ablation Study}
\label{sec:rq6}

We remove parts of \ourss~to see what each stage contributes, using Milan with the default private activities (Table~\ref{tab:ablation}). With only App-Conditioned Pre-training, the model generates realistic data but barely hides private activities (private accuracy 0.612). Adding App-Aware Fine-tuning without data filtering lowers private accuracy to 0.451 but lets low-quality samples drag non-private accuracy down to 0.798. Adding data filtering recovers non-private accuracy to 0.840 and pushes private accuracy to 0.387. Fine-tuning thus provides most of the protection, while filtering protects utility.

\begin{table}[t!]
\centering
\caption{Ablation on Milan. AAF provides most of the privacy protection, and data filtering protects non-private utility.}
\label{tab:ablation}
\resizebox{\columnwidth}{!}{

\begin{tabular}{lcc}
\toprule
\textbf{Configuration} & \textbf{Non-private Acc.} $\uparrow$ & \textbf{Private Acc.} $\downarrow$ \\
\midrule
ACP only                       & 0.831 & 0.612 \\
ACP + AAF (no filtering)       & 0.798 & 0.451 \\
ACP + AAF + filtering (\ourss) & \textbf{0.840} & \textbf{0.387} \\
\bottomrule
\end{tabular}
}
\end{table}












\section{Related Work}
\subsection{Data Privacy}
Data privacy has been studied from many angles, mainly through rule-based frameworks~\mbox{\cite{lola2023towards,aljeraisy2021privacy,gheisari2021obpp}} and differential-privacy frameworks~\mbox{\cite{husnoo2021differential,gao2022pfed,ghayyur2018iot}}. For rule-based work, Tamini et al.~\mbox{\cite{tamani2016towards}} built a user-centric strategy based on habit anomaly detection, and Lola et al.~\mbox{\cite{lola2023towards}} proposed a two-phase scheme where the manufacturer declares data-collection intent and the user declares preferences. These rely on humans to create and enforce rules, are error-prone, and manage streams individually rather than the cross-stream interactions central to our problem. For DP, Gao et al.~\mbox{\cite{gao2022pfed}} combined federated learning and LDP, but DP is less effective in multi-sensor settings because it ignores the relationships among sensors and degrades all applications regardless of privacy designation.
\subsection{Generative Models}
Generative models are widely used for data augmentation and have strong potential for the multi-sensor privacy challenge. GAN-based methods generate realistic time series: SensoryGANs~\mbox{\cite{wang2018sensorygans}} creates synthetic sensor data. TimeGAN~\mbox{\cite{yoon2019time}} and ActivityGAN~\mbox{\cite{li2020activitygan}} preserve temporal dynamics for HAR. However, GANs suffer from posterior and mode collapse, limiting the diversity of generated data. Diffusion models produce higher-quality streams; for example, NetDiffus~\mbox{\cite{sivaroopan2024netdiffus}} targets synthetic network traffic. Yet conditional generation in diffusion models~\mbox{\cite{batzolis2021conditional,zhang2023adding}} mostly targets images and is not suited to multi-sensor streams.



\section{Conclusion}

Multi-sensor environments enhance daily life, but their sensor fusion introduces application-level privacy risks. We propose \ourss, which uses contrastive learning within a diffusion model to generate privacy-preserving synthetic streams via two stages: App-Conditioned Pre-training (ACP) and App-Aware Fine-tuning (AAF). 
We also define a threat model for the multi-sensor sharing setting. Across three datasets spanning smart homes, a smart office, and an in-the-wild setting, \ourss~lowers private-application accuracy by 40 to 50\% while keeping non-private performance close to raw, outperforms both generative and simple baselines, and stays robust against an adaptive attacker.

\clearpage
\balance
\bibliographystyle{ACM-Reference-Format}
\bibliography{references}



\end{document}